\documentclass[a4paper,12pt]{article}
\usepackage{amsmath}
\usepackage{pstricks}
\usepackage{pst-node}
\usepackage[ansinew]{inputenc}
\usepackage{amssymb,amsmath}
\usepackage{amsfonts}
\usepackage{epsfig}

\def\p{\partial}
\def\px{\partial_x}
\def\py{\partial_y}
\def\a{\alpha}
\def\b{\beta}
\def\g{\gamma}
\def\d{\delta}
\def\o{\omega}

\font\Sets=msbm10

\def\Integer {\hbox{\Sets Z}}    

\def\Complex {\hbox{\Sets C}}   

\def\be{\begin{equation}}       \def\ba{\begin{array}}

\def\ee{\end{equation}}         \def\ea{\end{array}}

\def\bea {\begin{eqnarray}}      \def\eea {\end{eqnarray}}

\def\bean{\begin{eqnarray*}}    \def\eean{\end{eqnarray*}}

\def\pa  {\partial}

\def\const {\mathop{\rm const}\nolimits}

\def\RA {\ \Rightarrow\ }         

\def\qed   {\vrule height0.6em width0.3em depth0pt}

\def\<{\langle} \def\({\left(}  \def\>{\rangle} \def\){\right)}

\newtheorem{exi}{Example}

\author{E. Kartashova, A. Shabat}

\title{Computable Integrability.\\
Chapter 5: Factorization of LPDOs.}
\date{}
\begin{document}
%generates titel
\maketitle \tableofcontents {
\newpage
\section{Introduction}
Let us notice first that different definitions of integrability,
as a rule, use linearization of initial equation and/or expansion
on some basic functions which are themselves solutions of some
linear differential equation. Important fact here is that
linearization of some differential equation is its simplification
but not solving yet. For instance, in case of linear Schrödinger
equation, $\psi_{xx}+k^2\psi=u\psi$, we are not able  {\bf to
find} its solutions explicitly but only  {\bf to name} them  Jost
functions and to exploit their useful
properties (see previous Chapters).\\

On the other hand, well-known fact is that for LODE with constant
coefficients operator itself can always be factorized into
first-order factors and thus the problem is reduced to the solving
of a few first-order LODEs:
$$\frac{d}{dx}\psi + \lambda
\psi=f(x)$$
which are solvable in quadratures.\\

In case of differential operators with variable coefficients
factorization is not always possible but for the great number of
operators BK-factorization gives factorization conditions
explicitly which we are going to demonstrate in the next Section.

\section{BK-factorization}

Speaking generally, BK-factorization produces following result: in
case of LPDO with characteristic polynomial having at least one
distinct root, factorization is constructed algebraically for an
operator of arbitrary order $n$ while in case of some multiple
roots of characteristic polynomial of LPDO, factorization is
formulated in terms of Riccati equation(s). Factorization of  LODO
is always equivalent to solving some Riccati equation(s). Below
explicit procedure for order 2 and 3 is briefly described.

\subsection{LPDO of order 2}
Let us  outline here  BK-factorization procedure \cite{bk2005} for
the simplest case of bivariate LPDO of second order. Consider an
operator
\begin{equation}\label{A2}
A_2=\sum_{j+k\le2}a_{jk}\px^j\py^k
=a_{20}\px^2+a_{11}\px\py+a_{02}\py^2+a_{10}\px+a_{01}\py+a_{00}.
\end{equation}
with smooth coefficients and seek for  factorization
$$
A_2=(p_1\px+p_2\py+p_3)(p_4\px+p_5\py+p_6).
$$

Let us write down the equations on $p_i$ explicitly, keeping in
mind the rule of left composition, i.e. that $ \px (\alpha \py) =
\px (\alpha) \py +
\alpha \partial_{xy}.$\\

Then in all cases

$$
 \begin{cases}
  a_{20} &= \ p_1p_4\\
  a_{11} &= \ p_2p_4+p_1p_5\\
  a_{02} &= \ p_2p_5\\
  a_{10} &= \ \mathcal{L}(p_4) + p_3p_4+p_1p_6\\
  a_{01} &= \ \mathcal{L}(p_5) + p_3p_5+p_2p_6\\
  a_{00} &= \ \mathcal{L}(p_6) + p_3p_6
  \end{cases}
  \eqno (\it{2SysP})
 $$\\
where we use the notation $\mathcal{L} = p_1 \px + p_2 \py $. In
generic case we assume that (after a linear change of variables if
necessary)
$$
a_{20}\ne 0 \quad \mbox{and} \quad p_1=1.
$$

Then  first three equations of
 {\it 2SysP}, describing the highest order terms are
equations in the variables $p_2, p_4, p_5$ and to find them we
have to find roots of a quadratic polynomial
$$
\mathcal{P}_2(\o):=  a_{20}(\o)^2 +a_{11}(\o) +a_{02} = 0
$$

and this leads to a linear system for $p_4$, $p_5$ with $\o$ as
parameter:
$$
\begin{bmatrix} 1&0\cr -\o&1\end{bmatrix}\begin{bmatrix}p_4\cr p_5\end{bmatrix}
=\begin{bmatrix}a_{20}\cr a_{11}\end{bmatrix};\qquad
\begin{bmatrix}p_4\cr p_5\end{bmatrix}=
\begin{bmatrix}1&0\cr\o &1\end{bmatrix}
\begin{bmatrix}a_{20}\cr a_{11}\end{bmatrix}.
$$\\
Thus
 $$
  \begin{cases}
p_1=1\\
p_2=-\o\\
p_4=a_{20}\\
p_5=a_{20} \o +a_{11}
  \end{cases}
  \eqno (2{\it Pol})
 $$\\
and choice of a root $\o$ generates different possible
factorizations of operator $A_2$.\\

Having computed $p_2, p_4, p_5$ one can plug them into two next
equations of {\it 2SysP}
$$
  \begin{cases}
  a_{10} &= \ \mathcal{L}(p_4)+p_3p_4+p_1p_6\\
  a_{01} &= \ \mathcal{L}(p_5)+p_3p_5+p_2p_6
  \end{cases}\\
 $$
 and get a {\it linear} system of
equations in two variables $p_3,p_6$ which can easily be solved
\begin{eqnarray}
  p_3 =  \frac{\o a_{10}+a_{01} -\o\mathcal{L}a_{20}- \mathcal{L}(a_{20} \o+a_{11})}
{2a_{20}\o+a_{11}},\nonumber\\
  p_6 =\frac{ (a_{20}\o+a_{11})(a_{10}-\mathcal{L}a_{20})-a_{20}(a_{01}
 -\mathcal{L}(a_{20}\o+a_{11}))}{2a_{20}\o+a_{11}}.\nonumber\\
  \end{eqnarray}

if $\mathcal{P}_2'(\o)=2a_{20}\o+a_{11}\ne 0$, i.e. $\o$ is a
simple root. At this point all coefficients $p_1, p_2, ..., p_6$
have been computed
  and  condition of factorization

\be\label{cond2} a_{00} = \mathcal{L}(p_6)+p_3p_6 \ee

takes form\\
\be\label{cond2explicit}
\begin{cases}
a_{00} = \mathcal{L} \left\{
 \frac{\o a_{10}+a_{01} - \mathcal{L}(2a_{20} \o+a_{11})}
{2a_{20}\o+a_{11}}\right\}+ \frac{\o a_{10}+a_{01} -
\mathcal{L}(2a_{20} \o+a_{11})}
{2a_{20}\o+a_{11}}\times\\
\times\frac{ a_{20}(a_{01}-\mathcal{L}(a_{20}\o+a_{11}))+
(a_{20}\o+a_{11})(a_{10}-\mathcal{L}a_{20})}{2a_{20}\o+a_{11}}.
\end{cases}
 \ee

\subsection{LPDO of order 3}

Now we consider an operator
\begin{eqnarray}\label{A3}
A_3=\sum_{j+k\le3}a_{jk}\partial_x^j\partial_y^k =a_{30}\p_x^3 +
a_{21}\p_x^2 \py + a_{12}\px \py^2 +a_{03}\p y^3\\ \nonumber +
a_{20}\p_x^2+a_{11}\px\py+a_{02}\py^2+a_{10}\px+a_{01}\py+a_{00}.
\end{eqnarray}

with smooth coefficients and look for a factorization
$$
A_3=(p_1\px+p_2\py+p_3)(p_4 \p_x^2 +p_5 \px\py  + p_6 \py^2 + p_7
\px + p_8 \py + p_9).
$$

The conditions of factorization are described
 by the following system:
 $$
  \begin{cases}
  a_{30} &= \ p_1p_4\\
  a_{21} &= \ p_2p_4+p_1p_5\\
  a_{12} &= \ p_2p_5+p_1p_6\\
  a_{03} &= \ p_2p_6\\
  a_{20} &= \ \mathcal{L}(p_4)+p_3p_4+p_1p_7\\
  a_{11} &= \ \mathcal{L}(p_5)+p_3p_5+p_2p_7+p_1p_8\\
  a_{02} &= \ \mathcal{L}(p_6)+p_3p_6+p_2p_8\\
  a_{10} &= \ \mathcal{L}(p_7)+p_3p_7+p_1p_9\\
  a_{01} &= \ \mathcal{L}(p_8)+p_3p_8+p_2p_9\\
  a_{00} &= \ \mathcal{L}(p_9)+p_3p_9
   \end{cases}
  \eqno (\it{3SysP})
 $$
with $\mathcal{L} = p_1 \px + p_2 \py $.\\

Once again we may assume without loss of generality that the
coefficient of the term of highest order in $\p_x$ does not
vanish, and that the linear factor is normalized:
$$
a_{30}\ne 0,\qquad p_1=1.
$$
The first four equations of
 {\it 3SysVar} describing the highest order terms are
equations in the variables $p_2, p_4, p_5, p_6$.  Solving these
equations requires the choice of a root $-p_2$ of a certain
polynomial of third degree.  Once this choice has been made, the
remaining top order coefficients $p_4, p_5, p_6$ are easily found.
The top order coefficients can now be plugged into the next four
equations of {\it 3SysP}.  The first three of these four equations
will now be a {\it linear} system of equations in the variables
$p_3,p_7, p_8$ which is easily solved. The next equation is now a
{\it linear} equation on variable $p_9$ which means that all
variables $p_i, i=1,...,9$ have been found. The last two equations
of {\it 3SysP} will give us then the {\it conditions  of
factorization}.\\

 Namely, at {\bf the first step} from

$$
  \begin{cases}
  a_{30} = p_4 \\
  a_{21} = p_2p_4+p_5 \\
  a_{12} = p_2p_5+p_6\\
  a_{03} = p_2p_6
  \end{cases}\\
$$
it follows that
$$
\mathcal{P}_3(-p_2):=  a_{30}(-p_2)^3 +a_{21}(- p_2)^2 +
a_{12}(-p_2)+a_{03}=0.
$$
As for the case of second order, taking $p_2=-\o$, where $\o$ is a
root of the characteristic polynomial $\mathcal{P}_3$ we get a
linear system in $p_4, p_5, p_6$ with $\o$ as parameter.  Then
again
$$
p_2=-\o, \qquad
 \mathcal{L} = \px - \o \py,
$$
which leads to

$$
  \begin{cases}
  a_{30} = p_4 \\
  a_{21} = -\o p_4 + p_5 \\
  a_{12} = -\o p_5+p_6
  \end{cases}\\
$$
i.e.

$$
\begin{bmatrix} 1&0&0\cr -\o&1&0\cr 0&-\o&1\end{bmatrix}\begin{bmatrix}p_4\cr p_5\cr p_6\end{bmatrix}
=\begin{bmatrix}a_{30}\cr a_{21}\cr a_{12}\end{bmatrix};\qquad
\begin{bmatrix}p_4\cr p_5\cr p_6\end{bmatrix}=
\begin{bmatrix}1&0&0\cr\o &1&0 \cr \o^2& \o& 1 \end{bmatrix}
\begin{bmatrix}a_{30}\cr a_{21}\cr a_{12} \end{bmatrix}.
$$

Thus

 $$
  \begin{cases}
p_1=1\\
p_2=-\o\\
p_4=a_{30}\\
p_5=a_{30} \o+a_{21}\\
p_6=a_{30}\o^2+a_{21}\o+a_{12}.
  \end{cases}
  \eqno (3{\it Pol})
 $$

 At {\bf the second step}, from

$$
  \begin{cases}
  a_{20} &= \ \mathcal{L}(p_4)+p_3p_4+p_1p_7\\
  a_{11} &= \ \mathcal{L}(p_5)+p_3p_5+p_2p_7+p_1p_8\\
  a_{02} &= \ \mathcal{L}(p_6)+p_3p_6+p_2p_8\\
  \end{cases}\\
 $$
 and (2{\it Pol}) we get

 $$
  \begin{cases}
  a_{20}-\mathcal{L} a_{30} &=\ p_3 a_{30} +p_7\\
  a_{11}-\mathcal{L}(a_{30} \o+a_{21}) &=\ p_3(a_{30}\o+a_{21})- \o p_7+p_8\\
  a_{02}-\mathcal{L}(a_{30}\o^2+a_{21}\o+a_{12})&=\ p_3
(a_{30}\o^2+a_{21}\o+a_{12})-\o p_8.
  \end{cases}
  \eqno (3{\it Lin*})\\
 $$
As a linear system for $p_3$, $p_7$, $p_8$ this has determinant
$$
3a_{30}\o^2+2a_{21}\o+a_{12}=\mathcal{P}'_3(\o),
$$
so if $\o$ is a simple root the system has unique solution
\begin{eqnarray}
p_3 = \frac{\o^2 (a_{20} -\mathcal{L} a_{30})
+\o(a_{11}-\mathcal{L}(a_{30}\o+a_{21}))+a_{02}-\mathcal{L}
(a_{30}\o^2+a_{21}\o+a_{12})}{3a_{30}\o^2+2a_{21}\o+a_{12}};\nonumber\\
 p_7= \frac{a_{20}-\mathcal{L}a_{30}}{3a_{30}\o^2+2a_{21}\o+a_{12}}
-\frac{a_{30}}{3a_{30}\o^2+2a_{21}\o+a_{12}}\cdot p_3;\nonumber\\
 p_8= \frac{\o(a_{20}-\mathcal{L}a_{30})+a_{11}-\mathcal{L}(a_{30}\o+a_{21})}
{3a_{30}\o^2+2a_{21}\o+a_{12}} -\frac{a_{30}\o
+a_{21}}{3a_{30}\o^2+2a_{21}\o+a_{12}}\cdot p_3.\nonumber
\end{eqnarray}

 In order to find the last coefficient $p_9$ we use the next
equation of ({\it3SysVar}), namely:

$$
a_{10}  =  \mathcal{L}(p_7)+p_3p_7+p_1p_9, \eqno (3{\it Lin**}).
$$

At this point all coefficients $ p_i, i=1,...9$ have been
computed,
under the assumption that $\o$ is a simple root. \\

At {\bf the third step} from\\

\be \label{cond3}
\begin{cases}
a_{01} = \ \mathcal{L}(p_8)+p_3p_8+p_2p_9\\
a_{00} = \ \mathcal{L}(p_9)+p_3p_9
\end{cases}
\ee

 all the {\it necessary conditions} for factorization can be
written out.  We do not do so here because the formulas are
tedious and do not add anything to understanding the main idea. If
the conditions are satisfied,  the explicit factorization formulae
could be written out as for the second-order operator. The
difference is that in this case the polynomial defined by the
highest order terms is of degree 3
and we have not one but two conditions of factorization.\\

\subsection{Constant coefficients}
\begin{itemize}

\item{} Obviously, in case of constant coefficients $a_{ij}$ all
formulae above can be simplified considerably and used for
classical factorization problem of a polynomial. For instance, a
bivariate second order polynomial
$$
\mathcal{W}=X^2-Y^2+a_{10}X+a_{01}Y+a_{00}
$$
can be factorized into two linear polynomials,
$$
X^2-Y^2+a_{10}X+a_{01}Y+a_{00}=(p_1X+p_2Y+p_3)(p_4X+p_5Y+p_6),
$$
{\bf iff}
$$
a_{00}=\pm\frac{a_{01}^2-a_{10}^2}{4}.
$$

In each case coefficients $p_i$ can be written out explicitly, for
instance if
$$
a_{00}=\frac{a_{01}^2-a_{10}^2}{4},
$$
then
$$
X^2-Y^2+a_{10}X+a_{01}Y+a_{00}=(X+Y+\frac{a_{01}-a_{10}}{2})(X-Y+\frac{a_{01}+a_{10}}{2}).
$$

\item{}  As in case of order two,  constant coefficients $a_{ij}$
simplify all the formulae and reduce the problem under
consideration to the classical factorization  of a polynomial. For
instance, a bivariate third order polynomial
$$
\mathcal{W}=X^2Y+XY^2+a_{20}X^2+a_{11}XY+a_{02}Y^2+a_{10}X+a_{01}Y+a_{00}
$$
can be factorized into the product of one linear and one second
order polynomials,
$$
X^2Y+XY^2+a_{20}X^2+a_{11}XY+a_{02}Y^2+a_{10}X+a_{01}Y+a_{00}=$$$$=(p_1X+p_2Y+p_3)(p_4X^2+p_5XY+p_6Y^2+p_7
X + p_8 Y + p_9)
$$
for instance,  if
$$
a_{01}=a_{10}+(a_{20}+1)(a_{11}-a_{20}-a_{02}),$$$$
a_{00}=(a_{11}-a_{20}-a_{02})[a_{10}+a_{20}(a_{11}-a_{20}-a_{02})],
$$

and the result of factorization then is (with notation $\gamma=
a_{11}-a_{20}-a_{02}$):
$$
X^2Y+XY^2+a_{20}X^2+a_{11}XY+a_{02}Y^2+a_{10}X+a_{01}Y+a_{00}=$$
$$=(X+Y+\gamma)(XY-a_{20}X + (a_{20}-a_{11}+\gamma ) Y +
a_{10}+a_{20}\gamma).
$$
\end{itemize}

\section{Laplace transformation}
\subsection{Main notions}
The most important question now is - what to do when
conditions of factorization are violated? Do we still have a way
to solve an equation $\mathcal{L}(\psi)=0$ corresponding to the
initial operator? In order to answer these questions let us
re-write results of BK-factorization for generic case of second
order hyperbolic operator as

\begin{equation}\label{dar}
\mathcal{L}: \quad \p_x \p_y + a\p_x + b\p_y + c =
\left\{\begin{array}{c}
(\p_x + b)(\p_y + a) - ab - a_x + c\\
(\p_y + a)(\p_x + b) - ab - b_y + c
\end{array}\right.
\end{equation}

and corresponding LPDE as $(\p_x \p_y + a\p_x + b\p_y + c)\psi_1
=0$ and introduce new function $\psi_2= (\p_y + a)\psi_1$. Our
main goal now is to construct some {\bf new} LPDE having $\psi_2$
as a solution and to check its factorization property. If this new
LPDE is factorizable, then its solution is written out explicitly
and due to the invertibility of a transformation $\psi_1
\rightarrow \psi_2$ the
formula for solution of initial LPDE can also be obtained immediately. \\

Let us first introduce some definitions.

\paragraph{Definition 3.1 } Two operators of order $n$
$$
\mathcal{L}_1=\sum_{j+k\le n}a_{jk}\partial_x^j\partial_y^k \quad
\mbox{and} \quad \mathcal{L}_2=\sum_{j+k\le
n}b_{jk}\partial_x^j\partial_y^k
$$
are called {\it equivalent operators} if  there exists some
function $f=f(x,y)$ such that
$$
f \mathcal{L}_1= \mathcal{L}_2 \circ f.
$$
\paragraph{Definition 3.2 } Expressions $$\hat{a}=  ab  +a_x - c \quad
\mbox{and} \quad \hat{b}=  ab  +b_y - c$$ are called {\it Laplace
invariants}.

\paragraph{Lemma 3.3 } Two hyperbolic operators $\mathcal{L}_1$ and
$\mathcal{L}_2$ of the form (\ref{dar}) are equivalent {\bf iff}
their Laplace invariants coincide
pairwise.\\

$\blacktriangleright$ Indeed,
$$
f\mathcal{L}_1(\psi)=f\p_x \p_y + fa_1\p_x + fb_1\p_y + fc_1
$$
and
$$
\mathcal{L}_2(f\psi )=\p_x \p_y(f )\psi + \p_yf \p_x \psi + \p_x
\p_y (\psi )f + \p_y \psi \p_x f+$$$$
 a_2\p_x( f) \psi + a_2\p_x (\psi) f+
b_2\p_y (f) \psi + b_2\p_y (\psi) f + f\psi c_2,
$$
i.e. $ f \mathcal{L}_1= \mathcal{L}_2 \circ f $ iff
$$
a_1=a_2+ \frac{\p_yf}{f}=a_2+ (\log f)_y;
$$
$$
b_1=b_2+ \frac{\p_xf}{f}=b_2+ (\log f)_x;
$$
$$
c_1=c_2+ \frac{\p_x\p_yf}{f}+a_2\frac{\px f}{f} + b_2\frac{\py
f}{f}=c_2+(\log f)_{xy}+ (\log f)_x (\log f)_y+$$$$+ a_2(\log
f)_x+b_2 (\log f)_y.
$$

Direct substitution of these expressions into formulae for Laplace
invariants gives (we use notation $\varphi = \log f$):

$$
\hat{a}_1=a_1b_1 + a_{1,x} - c_1=(a_2+
\varphi_y)(b_2+\varphi_x)+(a_2+ \varphi_y)_x -
$$
$$
c_2-\varphi_{xy}- \varphi_x \varphi_y- a_2\varphi_x-b_2 \varphi_y=
a_2b_2+ a_{2,x}-c_2= \hat{a}_2.
$$

Analogously one can obtain $\hat{b}_1=\hat{b}_2$ and it means that
for two equivalent hyperbolic operators their Laplace invariants
do coincide.\qed \\

$\blacktriangleright$ First of all, let us notice that  two
operators
$$
 \mathcal{L}_1= (\p_x + b_1)(\p_y + a_1)+A_1
$$
and
$$
 \mathcal{L}_2= (\p_x + b_2)(\p_y + a_2)+A_2
$$
can be transformed into some equivalent form
$$
 \tilde{\mathcal{L}}_1= (\p_x + b_1)\p_y + A_1
$$
and
$$
 \tilde{\mathcal{L}}_2= (\p_x + b_2)\p_y + A_2
$$
(perhaps) by different functions $f_1, \ f_2$ such that
$$
f_1 \mathcal{L}_1=  \tilde{\mathcal{L}}_1 \circ f_1, \ \ f_1
\mathcal{L}_2=  \tilde{\mathcal{L}}_2 \circ f_2
$$

and as was proven above, Laplace invariants of the initial
operators coincide with those of the equivalent ones. Operator of
the form
$$
\mathcal{L}= (\p_x + b)\p_y + A
$$
has following Laplace invariants: $ A$ and  $-b_y +A$, i.e. $
A_1=A_2 $ and $b_{1y}=b_{2y} $  for operators $\mathcal{L}_1$ and
$\mathcal{L}_2$ with the same Laplace invariants. It yields to
$$
b_1=b_2 + \varphi (x), \ \   \tilde{\mathcal{L}}_2= (\p_x + b_1 +
\varphi (x))\p_y + A_1
$$
with some arbitrary smooth function $\varphi (x)$. Now operator
$\tilde{\mathcal{L}}_2$ differs from $\tilde{\mathcal{L}}_1$ only
by term $\varphi (x)$ which can be "killed" by one more equivalent
transformation, namely, for some function $f_3=f_3(x)$
$$
\tilde{\tilde{\mathcal{L}}}_2= f_3^{-1} \tilde{\mathcal{L}}_2
\circ f_3=(\p_x + b_1 + \varphi (x)+ (\log f_3)_x)\p_y + A_1
$$
and choice  $\varphi (x)=f_3^{'}/f$ completes the proof.
 \qed \\

Now let us rewrite  initial operator (\ref{dar}) as
$$
\mathcal{L}_1: \quad \p_x \p_y + a_1\p_x + b_1\p_y + c_1
$$

and notice that
$$(\p_x + b_1)\psi_2=\hat{a}_1\psi_1 $$ leads to
$$
\left\{\begin{array}{c}
(\p_x + b_1)\psi_2= \hat{a}_1\psi_1 \\
  (\p_y + a_1)\psi_1=\psi_2
\end{array}\right.
\RA \left\{\begin{array}{c}
\frac{1}{\hat{a}_1}(\p_x + b_1)\psi_2= \psi_1 \\
  (\p_y + a_1)\psi_1=\psi_2
\end{array}\right.
 $$

 and standard formula for  {\bf log-derivative}:
 $$ e^{\varphi}\p_y  e^{-\varphi}= \p_y -
\varphi_y \quad \mbox{with} \quad \varphi=\log{\hat{a}},$$
 gives finally
a new operator $\mathcal{L}_2$ with corresponding LPDE
\begin{equation}\label{hatL}
\mathcal{L}_2(\psi_2)\equiv \Big[\Big(\p_y + a_1 -
(\log{\hat{a}_1})_y \Big)(\p_x +
b_1)-\hat{a}_1\Big]\psi_2=0.\end{equation}

In order to check whether these two operators $\mathcal{L}_1$ and
$\mathcal{L}_2$ are different, let us compute  Laplace
invariants of the new operator $\mathcal{L}_2$:
$$
\begin{cases}
\hat{a}_1=  a_1b_1  +a_{1,x} - c_1 \\
\hat{b}_1=  a_1b_1  +b_{1,y} - c_1
\end{cases}
\RA
\begin{cases}
\hat{a}_2= \hat{a}_1- \hat{a}_{1,y}- (\log \hat{a}_1)_{xy}+a_{1,x} \\
\hat{b}_2=  \hat{a}_1
\end{cases}
$$
Now one can see that operators   $\mathcal{L}_1$ and
$\mathcal{L}_2$ are  not equivalent and  operator $\mathcal{L}_2$
is factorizable if $\hat{a}_2 =0$ or
$\hat{b}_2=0$ (see example below).\\

\paragraph{Definition 3.4}
Transformation $\mathcal{L}_1 \RA \mathcal{L}_2$, i.e.
$$
\begin{cases}
\hat{a}_1=  a_1b_1  +a_{1,x} - c_1 \\
\hat{b}_1=  a_1b_1  +b_{1,y} - c_1
\end{cases}
\RA
\begin{cases}
\hat{a}_2= \hat{a}_1- \hat{a}_{1,y}- (\log \hat{a}_1)_{xy}+a_{1,x} \\
\hat{b}_2=  \hat{a}_1
\end{cases}
$$
is called {\it Laplace transformation.}\\

If first new operator is also not factorizable, the procedure can
be carried out for as many steps as necessary in order to get some
factorizable operator.  At the step $N$ when the first
factorizable operator is found, algorithm stops because the
division on corresponding $\hat{a}_N =0$ is not possible any
more. In fact, it is possible to write out formulae for Laplace transformation in terms of
Laplace invariants only.

\paragraph{Theorem 3.5 } Let $u_n$ is one of Laplace invariants $\hat{a}_n,
\hat{b}_n$ obtained at the step $n$. Then

\be \label{nLaplace}u_{n+1}=2u_n+(\log{u_n})_{xy}-u_{n-1}. \ee

$\blacktriangleright$ Indeed, due to Lemma 3.3 it is enough to regards sequence
of operators of the form
$$
\mathcal{L}_n: \quad \p_x \p_y + a_n\p_x + b_n\p_y + c_n \quad
\mbox{with } a_n=0
$$
because it is easy to find some function $f_{a_n}$ (for instance,
$f_{a_n}=e^{-\int a_ndy}$) such that

\be \label{A0}\mathcal{A}_{n,0}=f_{a_n}^{-1}\mathcal{L}_n \circ
f_{a_n} = \p_x \p_y + \tilde{b}_n\p_y + \tilde{c}_n. \ee

From now on
$$
\mathcal{L}_n: \quad \p_x \p_y + b_n\p_y + c_n
$$
and tilde-s are omitted for simplicity of notations. Now formulae
for Laplace transformation take form

\be \label{LaplaceTrans}\psi_{n,y}=-c_n\psi_{n+1}, \quad
\psi_n=[\px+b_n+(\log c_n)_x]\psi_{n+1} \ee

and Eq.(\ref{hatL}) can be rewritten for the function $\psi_{n+1}$
as
$$
[\px+b_n+(\log {c_n})_x]\py
\psi_{n+1}+[c_n+b_{n,y}+(\log{c_n})_{xy}]\psi_{n+1}=0,
$$
i.e.
$$
c_{n+1}-c_n=b_{n,y}+(\log{c_n})_{xy}, \quad b_{n+1}-b_n=
(\log{c_n})_{x},
$$
$$
c_{n+2}-c_{n+1}=b_{n+1,y}+(\log{c_{n+1}})_{xy}, \quad
b_{n+2}-b_{n+1}= (\log{c_{n+1}})_{x}
$$
and finally \be
\label{TodaC}c_{n+2}=2c_{n+1}+(\log{c_{n+1}})_{xy}-c_n. \ee Notice
that in this case, first Laplace invariant is
$$\hat{a}_n=  a_nb_n  +a_{n,x} - c_n =-c_n $$
and obviously satisfies to Eq.(\ref{TodaC}), i.e. for the first
invariant the statement of the theorem is proven. In order to
prove it for the second invariant $\hat{b}_n$ one has to choose
another sequence of operators with $b_n=0$ generated by some
function $f_{b_n}$ such that

\be \label{B0}\mathcal{B}_{n,0}=f_{b_n}^{-1}\mathcal{L}_n \circ
f_{b_n} = \p_x \p_y + \tilde{a}_n\p_x + \tilde{c}_n. \ee
 \qed

Notice that in order to obtain the recurrent formula for Laplace
invariants, we used separation of variables $x$ and $y$ given by
(\ref{LaplaceTrans}). Moreover, introduction of a new discrete
variable $n$ allows us to regards these equations as
difference-differential ones. In order to deal with this sort of
equations one needs a couple of definitions.

\paragraph{Definition 3.6 } An operator $T$ acting on the
infinite sequences of functions
$$(...,\psi_{-2},
\psi_{-1}, \psi_0, \psi_1, \psi_2 ,... ,\psi_n,...)$$
 as $$ T\psi_n=\psi_{n+1}, \quad T^{-1}\psi_n=\psi_{n-1}$$
is called {\bf shift operator}. For convenience of notation
sometimes infinite vector-function $ \vec{\psi}_{\infty}$ is
introduced
$$
\vec{\psi}_{\infty}=(...,\psi_{-2}, \psi_{-1}, \psi_0, \psi_1,
\psi_2 ,... ,\psi_n,...)
$$
and  matrix of operator $T$ then has the following form:

\be\label{T}   T=
 \left(
 \begin{array}{llllllll}
... & 0 &  1  & 0 & 0 & \dots  & 0 & ...\\
... & 0 &  0  & 1 & 0 &\dots  &  0 & ... \\
 ... & 0 & 0 & 0 & 1 & \dots  &  0  & ...\\
 ... & && \ddots  & \ddots & \ddots  &  & ...\\
 ... & 0 & 0 & \dots & 0 & 0 & 1 & ...\\
... & 0 &   0     & \dots & 0 & 0 &  0& ...
\end{array}
\right), \ee i.e. it is infinite matrix with all zero elements but
the elements over main diagonal - they are equal to 1.

\paragraph{Definition 3.7 } Commutator $\mathcal{C}=[A,B]$ of two operators $A$ and $B$
is defined as $$\mathcal{C}=AB-BA.$$\\

Obviously, following properties hold true:
\begin{itemize}
\item{} $[\px,\py]=0$ (cross-derivative rule), \item{}
$[\px,T]=[\py,T]=0$, \item{}$[T,a]=T\circ a-a \circ T=
(T(a)-a)\circ T$ (Leibnitz rule analog).
\end{itemize}

Let us now regard two operators corresponding
 Laplace transformations from Theorem 3.5 rewriting slightly
formulae (\ref{LaplaceTrans}) in terms of shift operator: \be
\label{LAXb}\psi_{n,y}=-c_nT\psi_n, \quad \psi_{n,x}+b_n\psi_{n}=
T^{-1}\psi_n.\ee

The use of shift operator makes it possible to present their
commutator
$$
\mathcal{C}=[\py+c_nT, \ \ \px+b_n-T^{-1}] \equiv [\py+cT,\ \
\px+b-T^{-1}]
$$
omitting low index $n$.
\paragraph{Lemma 3.8 }
Commutator  $$\mathcal{C}=[\py+cT,\ \ \px+b-T^{-1}]$$ is equal to
zero {\bf iff}

 \be \label{TodaB}
c_{x}=c(T(b)-b), \quad b_{y}=c -T^{-1}(c). \ee

$\blacktriangleright$ Indeed, by definition
$$
\mathcal{C}=b_y-c_xT+cTb-bcT+T^{-1}cT-c=-c+b_y+T^{-1}(c)-(c_x-cT(b)+bc)T.
$$
Now, if $ \mathcal{C}=0$, then $-c+b_y+T^{-1}(c)=0$ and
$c_x-cT(b)+bc=0$, i.e. (\ref{TodaB}) holds.

If (\ref{TodaB}) holds, then coefficients of $ \mathcal{C}$ are
equal to zero, i.e. $ \mathcal{C}=0$. \qed \\

 At the end of this Section let us notice that in the original Eq.(\ref{dar}) two variables $x$ and
$y$ played symmetrical role which can also be observed in
commutation relation of Lemma 3.8 after appropriate gauge
transformation:
$$
e^{q_n}(\py+c_nT)e^{-q_n}=\py-a_n+T, \ \
e^{q_n}(\px+b_n+T^{-1})e^{-q_n}=\px+c_{n-1}T^{-1}
$$
with
$$
c_n=e^{q_{n+1}-q_n},\  \ b_n=q_{nx}, \ \  a_n=q_{ny}.
$$
It can be shown that $q_n$ satisfy the following equation \be
\label{toda} q_{n,xy}=e^{q_{n+1}-q_n}-e^{q_{n}-q_{n-1}}.\ee

The equation (\ref{toda}) is usually called {\bf two-dimensional
Toda lattice} and plays fundamental role in the theory of Laplace
transformations.

\subsection{Truncation condition}

\paragraph{Definition 3.9 } Truncation condition for
Eq.(\ref{nLaplace}), namely
$$(\log{u_n})_{xy}=u_{n+1}-2u_n+u_{n-1}, \quad n=1,...,N+1$$
is defined by Dirichlet boundary conditions, i.e. $u_0=
u_{N+1}=0.$\\

For example, case $\boxed{N=1}$ gives us {\bf Liouville
equation}{\footnote{See Ex.1}} $$(\log u_1)_{xy}+2u_1=0$$ while
case $\boxed{N=2}$ yields to the system
$$
\begin{cases}
(\log{u_1})_{xy}=u_{2}-2u_1+u_{0}\\
(\log{u_2})_{xy}=u_{3}-2u_2+u_{1}
\end{cases} \RA
\begin{cases}
(\log{u_1})_{xy}  =  -2u_1+u_{2}\\
(\log{u_2})_{xy}  = \ \ u_{1}-2u_2
\end{cases}
$$
These both cases are known to be integrable in quadratures. \\

Case of arbitrary given $N$ corresponds to the system of equations
with
 following matrix

 \be\label{AN}   A_N=
 \left(
 \begin{array}{llllll}
 -2 &  1  & 0 & 0 & \dots  & 0\\
1 &  -2  & 1 & 0 &\dots  &  0  \\
 0 & 1 & -2 & 1 & \dots  &  0  \\
 && \ddots  & \ddots & \ddots  &  \\
 0 & 0 & \dots & 1 & -2 & 1 \\
0 &   0     & \dots & 0 & 1 &  -2
\end{array}
\right) \ee

on the right hand. This matrix is some {\bf Cartan matrix} and
another choice of boundary conditions leads to  Cartan matrices of
other form. It is interesting to notice that {\bf all} Cartan
matrices can be constructed in this way. Moreover,
 it is proven that  system of equations corresponding to each Cartan matrix is
integrable in quadratures (\cite{lez}). Most well-known source of
Cartan matrices is semi-simple classification of algebras Lie -
there is exists one-to-one correspondence between these matrices
and semi-simple algebras Lie (\cite{lez}).\\

Explicit expression for any Laplace invariant $u_n$ after $n$
Laplace transformations is given by following lemma (\cite{lez})
which for simplicity is formulated for the special case Toda
chain.

\paragraph{Lemma 3.10 } Let us regard an infinite sequence of functions
$$\{d_n\}, \ n=0,1,2,...$$ such that
$$\px\py \log{d_n}=\frac{d_{n+1}d_{n-1}}{d^2_{n}}, \ \ \forall n$$
and
$$d_0=1, \ \ d_1=w(x,y)$$
for some smooth function $w(x,y)$ of two variables $x,y$. Then
$$
d_n=\det (\px^i\py^j w), \ i,j=1,...,n.
$$
\paragraph{Corollary 3.11 }
Sequence of functions
$$ u_n=\frac{d_{n+1}d_{n-1}}{d^2_{n}}$$
is solution of  Eq.(\ref{nLaplace}) while sequence of functions
$$e^{q_n}=\frac{d_n}{d_{n-1}}$$
is solution of Eq. (\ref{toda}).\\

Theorem 3.5  describes an infinite chain of equations
corresponding to Laplace transformations and to start with this
chain, we need nothing more then two invariants. On the other
hand, many applications of this theorem are connected with some
special problems in which different sort of finite chains are
considered. In the next Section we will  discuss  two most usable
ways to construct some finite chain of invariants and we close
this Section with an example of 2-steps chain.

\paragraph{Example 3.12 }

Let us regard operator
$$
\mathcal{L}_1= \p_x \p_y + x\p_x  + 2,
$$
then its Laplace transformation gives
$$
\begin{cases}
\hat{a}_1=  -1 \\
\hat{b}_1=  -2
\end{cases}
\RA
\begin{cases}
\hat{a}_2= 0 \\
\hat{b}_2=  -1
\end{cases}
$$
and operator $\mathcal{L}_2$ is factorizable:
$$
\mathcal{L}_2=\px\p_y+x\px+1=\px(\py+x).
$$
It is a simple task to write out explicitly solution $\psi_2$ of
LPDE \be \label{step1} \mathcal{L}_2(\psi_2):=\px(\py+x)(\psi_2)=0
\ee and afterwards solution $\psi_1$ of
$$ \mathcal{L}_1(\psi_1):= (\p_x \p_y + x\p_x  + 2)(\psi_1)=0 $$
can be computed by formula \be \label{end1}\psi_1= -\p_x
\psi_2.\ee

Indeed, introducing in Eq.(\ref{step1}) notation
$\varphi=(\py+x)\psi_2$ we find that $\varphi=Y(y)$ is arbitrary
function of one variable (...)
$$
\psi_2= X(x)e^{-xy}+ \int e^{x(y'-y)}Y(y')dy'.
$$

\vspace{5mm}

\subsection{Periodic closure}

\paragraph{Definition 3.13 } {\bf Classical periodic closure} for
equation on Laplace invariants
$$(\log{u_n})_{xy}=u_{n+1}-2u_n+u_{n-1}, \quad n=1,...,N$$
is defined by periodic boundary conditions, i.e. $u_{n+N}=
u_{n}.$\\

In this case Cartan matrix $A_N$ is replaced by matrix
$\tilde{A}_N$ and for $N \ge 3$ its form is

 \be\label{tildeAN}\tilde{A}_N=
 \left(
 \begin{array}{llllll}
 -2 &  1  & 0 & 0 & \dots  & 1\\
1 &  -2  & 1 & 0 &\dots  &  0  \\
 0 & 1 & -2 & 1 & \dots  &  0  \\
 && \ddots  & \ddots & \ddots  &  \\
 0 & 0 & \dots & 1 & -2 & 1 \\
1 &   0     & \dots & 0 & 1 &  -2
\end{array}
\right) \ee

Notice{\footnote{Ex.2}} that matrix $\tilde{A}_N$ {\bf is
degenerated}. It will be shown below that for $N=1$ initial equation can be solved explicitly,
while for $N=2,3$ initial system of equations allows some reduction to {\bf one} scalar equation. \\

Let us regard first case $\boxed{N=1}$ , it yields to
$(\log{u_1})_{xy}=u_{2}-2u_1+u_{0}=0$ and obviously $u_1=
g_1(x)g_2(y)$ with  arbitrary smooth functions $g_1(x),g_2(y)$.\\

Case $\boxed{N=2}$ is more interesting due to the huge amount of
applications (surfaces with constant curvature, relativity theory,
etc. ) and gives rise to  the system of equations

\be \label{z2} \begin{cases}(\log{u_1})_{xy}=2(u_{1}-u_2),\\
(\log{u_2})_{xy}=2(u_{2}-u_{1})\end{cases}\ee

with many important properties: it has conservation laws,
symmetries, soliton-type particular solutions, etc. In particular,
the reduction of this system can easily be constructed to the one
scalar equation:
$$
(\log{u_1})_{xy}+(\log{u_2})_{xy}=0 \RA (\log{u_1 u_2})_{xy}=0 \RA
u_1 u_2 = X(x) Y(y)
$$
with two smooth arbitrary functions $X(x), Y(y)$. Suppose now $u_1
u_2=1$, then setting $u_1=e^\theta, \ \ u_2=e^{-\theta}$ we find
solutions of Sys.(\ref{z2}) from solutions of equation
 \be \label{sinh} \theta_{xy}+\sinh \theta=0. \ee

This equation is called {\bf sinh-Gordon equation} \cite{bianchi}.\\

Case $\boxed{N=3}$ corresponds to the system
$$
\begin{cases}
(\log{u_1})_{xy}=u_2+u_3-2u_1\\
(\log{u_2})_{xy}=u_1+u_3-2u_2\\
(\log{u_3})_{xy}=u_1+u_2-2u_3
\end{cases}
$$
which allows reduction $ u_1=u_3$
$$
\begin{cases}
(\log{u_1})_{xy}=u_2+u_1-2u_1\\
(\log{u_2})_{xy}=u_1+u_1-2u_2\\
(\log{u_1})_{xy}=u_1+u_2-2u_1
\end{cases}
\RA
\begin{cases}
(\log{u_1})_{xy}=u_2-u_1\\
(\log{u_2})_{xy}=2(u_1-u_2)
\end{cases}
$$
and it can be treated analogously with the case above:
$$u_1u_2u_3=1, \ \ u_2=e^{-2\theta}, \ \ u_1=e^{\theta}, \ \ u_3=
e^{\theta} \RA$$
 \be \label{z3}
\theta_{xy}=e^{-2\theta}-e^{\theta}. \ee

Eq.(\ref{z3}) is called {\bf Tzitzeica equation} and  its
solutions give solution of initial system of equations. Tzitzeica
equation is also
very important for various applications \cite{ziz}.\\

For both Eq.(\ref{sinh}) and Eq.(\ref{z3}) their general solutions
{\bf are not available} as well as for the case of general $N$.
Attempts to solve appearing systems of equations directly demand
some tedious technique of inverse scattering
 and produce partial solutions with singularities.
 Method to reduce the initial system to one equation
simplifies drastically construction of smooth solutions. \\

At the end of this section let us notice that the same three
equations which were obtained while studying  truncated and
periodical cases - namely, Liouville, sinh-Gordon and Tzizeica
equations - do appear together in some other context \cite{sha1}:

\paragraph{Theorem.} Nonlinear PDE of the form $u_{xy}=f(u)$ has higher symmetries
{\bf iff} one of three cases take place: \be \label{3func}f=e^u,\
\ f=e^u+e^{-u}, \ \ f=e^u+e^{-2u}.
\ee\\

As we know already (see Chapter 1) that integrability of a
differential equation is intrinsically related to its symmetry
properties. Of course, possession  of a symmetry {\bf does not
mean} that equation is integrable in some sense but this fact
gives us a good hint on what equations {\bf might be integrable}.
Moreover, for some classes of differential  equations it is proven
\cite{sha2} that integrability is equivalent to some well-defined
symmetry properties.\\

 From this point of view the theorem above justifies
  hypothesis that two of these three equations (integrability in quadratures
  of Liouville equation we have demonstrated already) have good integrability properties.
  As we will show in the next section,  integrability of these PDEs
can be reduced to the integrability of well-know ODEs.
\\

 Thus, in
contrast to the truncation condition which leads to integrability
in quadratures for arbitrary order $N$, periodical closure
generates a more complicated situation. Here we are not able to
get the answer directly in terms of invariants but we need first
to study the properties of solutions $\psi_n$ at the all $N$ steps
simultaneously.

\subsection{Separation of variables}

Obviously, in case of periodical closure with period $N$ functions
$\psi_1$ and $\psi_{N+1}$ satisfy the same equation which was
presented in Theorem 3.6:

\be \label{psina} \psi_{n,xy}+b_{n}\psi_{n,y}+c_{n}\psi_{n}=0, \ \
\forall \ \ n=0, \pm1, \pm2,... \ee

or, in matrix form,
$$
(\px\py+\{b\}\py+\{c\})\vec{\psi}_{\infty}=0
$$
with diagonal  matrices $\{b\}$ and $\{c\}$. This fact allows us
to regard {\bf finite} vector-function $\vec{\psi}$,
$$
\vec{\psi}=(\psi_1,...\psi_{N}),
$$
which defines completely all the properties of the initial
infinite system of equations. Notice that the fact of coincidence
for coefficients of two LPDEs {\bf does not mean} that their
solutions $\psi_{n+N}$ and $\psi_{n}$ also do coincide - they
might differ, for instance, by a constant multiplier. Therefore we
need now some notion of periodic closure for solutions.

\paragraph{Definition 3.14 } {\bf Bloch periodic closure} is
defined for the components of function $ \vec{\psi}$ as follows:
$$
\psi_{n+N}=k^N\psi_{n}.
$$

Notice once more that truncation and classical periodical closure
are defined for Laplace invariants, i.e. for {\bf coefficients} of
Eqs.(\ref{psina}), while Bloch closure deals with {\bf solutions}
of the same equations.

\paragraph{Corollary 3.15 }  Shift
matrix $T_N$ in case of Bloch periodical closure has form

\be\label{T_N}   T_N=
 \left(
 \begin{array}{llllll}
 0 &  1  & 0 & 0 & \dots  & 0\\
0 &   0  & 1 & 0 &\dots  &  0  \\
 0 & 0 &  0 & 1 & \dots  &  0  \\
 && \ddots  & \ddots & \ddots  &  \\
 0 & 0 & \dots & 0 &  0 & 1 \\
k^N &   0     & \dots & 0 & 0 &   0
\end{array}
\right) \ee where $k \in \Complex$ is a free parameter.

\paragraph{Corollary 3.16 }
 Basic Lemma 3.8 holds true also for the
case of periodic closure, i.e. commutator of Laplace
transformations $$\mathcal{C}=[\py+cT_N,\ \ \px+b-T_N^{-1}] = 0$$
{\bf iff}
$$
c_{n,x}=c_n(b_{n+1}-b_{n}), \quad b_{n,y}=c_n -c_{n-1}, \quad n
\in \Integer_N,$$ which can be checked directly.

\paragraph{Example 3.17 }

As it was shown above, in case of $N=2$ which corresponds to the
classical periodic closure of $u_n$, chain of invariants
degenerates into sinh-Gordon equation  and this closure has the
form (\ref{z2}). Corollary 3.14 allows us to construct a
connection between closure
for invariants and closure for solutions of(\ref{LaplaceTrans}) :\\

$$
\begin{bmatrix}\psi_1\cr \psi_2\end{bmatrix}_x=
\begin{bmatrix}-b_1 & k^{-2} \cr 1 & -b_2\end{bmatrix}
\begin{bmatrix}\psi_1\cr \psi_2 \end{bmatrix}, \quad
\begin{bmatrix}\psi_1\cr \psi_2\end{bmatrix}_y=
\begin{bmatrix}0 & -c_1 \cr -k^2c_2 & 0\end{bmatrix}
\begin{bmatrix}\psi_1\cr \psi_2 \end{bmatrix}.
$$
Then
$$
\psi_{1x}+b_1\psi_1=k^{-2}\psi_2, \ \
\psi_{1}=b_2\psi_2+\psi_{2x},
$$
and excluding  $\psi_1$ or $\psi_2 $ we get an equation of the
second order on {\bf one} scalar function, for instance
$$
\psi_{2,xx}+(b_1+b_2)\psi_{2x}+(b_{2x}+b_1b_2-k^{-2})\psi_2=0.
$$
This equation is obviously equivalent to linear Shrödinger
equation
$$
\psi_{xx}=(\lambda + u)\psi, \quad \lambda= k^{-2},
$$
i.e. Sinh-Gordon equation is S-integrable. This important fact
plays role, for instance, while constructing surfaces of constant
curvatures (see very exhaustive review \cite{bob}).\\

\section{General invariants and semi-invariants}

Before discussing the notion of general invariant, let us notice
that  {\bf arbitrary} LPDO of second order, $A_2$,  can be
represented in the form  of factorization with reminder
$$
 A_2
=(p_1\px+p_2\py+p_3)(p_4\px+p_5\py+p_6)-l_2
$$
where reminder $l_2$ is defined by (\ref{cond2}):

\be\label{l2} l_2=  a_{00} - \mathcal{L} \left\{
 \frac{\o a_{10}+a_{01} - \mathcal{L}(2a_{20} \o+a_{11})}
{2a_{20}\o+a_{11}}\right\}- \frac{\o a_{10}+a_{01} -
\mathcal{L}(2a_{20} \o+a_{11})} {2a_{20}\o+a_{11}}\times$$$$
\times\frac{ a_{20}(a_{01}-\mathcal{L}(a_{20}\o+a_{11}))+
(a_{20}\o+a_{11})(a_{10}-\mathcal{L}a_{20})}{2a_{20}\o+a_{11}}.
 \ee

Similar to the case of order two,
 arbitrary LPDO of third order $A_3$ can be represented
 in the following form
$$ A_{3}=(p_1\px+p_2\py+p_3)(p_4 \p_x^2 +p_5 \px\py  + p_6 \py^2 +
p_7 \px + p_8 \py + p_9)-l_3\py-l_{31}. $$ In contrast to the
second order LPDO, in this case factorization with reminder gives
us not a function but a linear first order operator and it is
convenient for our further investigations to regard in this case
two "reminders" $l_3$ and $l_{31}$ which are defined by
Sys.(\ref{cond3}):
$$l_3= a_{01}- (p_1\px+p_2\py+p_3)p_8-p_2p_9
$$
$$ l_{31}=a_{00}-(p_1\px+p_2\py+p_3)p_9.$$

In the next Section it will be shown that  "reminders" $l_2, \
l_3, \ l_{31}$ are invariants of the corresponding operators uner
the equivalence transformations.

\subsection{Construction of invariants}
Let us first recollect  definition of two equivalent operators.

\paragraph{Definition 4.1 } Two operators of order $n$
$$
\mathcal{L}_1=\sum_{j+k\le n}a_{jk}\partial_x^j\partial_y^k \quad
\mbox{and} \quad \mathcal{L}_2=\sum_{j+k\le
n}b_{jk}\partial_x^j\partial_y^k
$$
are called {\it equivalent operators} if  there exists some
function $f=f(x,y)$ such that
$$
f \mathcal{L}_1= \mathcal{L}_2 \circ f.
$$

The definition is given for an operator  $ \mathcal{L}$ of
arbitrary order $n$ and obviously {\bf any} factorization
$$
\mathcal{L}=\mathcal{L}_1\mathcal{L}_2
$$
can be written out in equivalent form
$$
f^{-1} \mathcal{L}\circ f=f^{-1}\mathcal{L}_1\mathcal{L}_2\circ
f=(f^{-1}\mathcal{L}_1f)(f^{-1}\mathcal{L}_2\circ f),
$$
as well as sum of operators
$\mathcal{L}=\mathcal{L}_1+\mathcal{L}_2$:
$$
f^{-1} \mathcal{L}\circ f=f^{-1}(\mathcal{L}_1+\mathcal{L}_2)\circ
f=(f^{-1}\mathcal{L}_1\circ f)+(f^{-1}\mathcal{L}_2\circ f).
$$
Below we will take function $f$  in a form $f=e^{\varphi}$ for
convenience. In order to formulate theorem on invariants we need
following notations:

$$ A_{2a}=
a_{20}\p_x^2+a_{11}\px\py+a_{02}\py^2+a_{10}\px+a_{01}\py+a_{00},
$$

$$ A_{2p}=(p_1\px+p_2\py+p_3)(p_4 \p_x +p_5 \py+
p_6)-l_2, $$

$$ A_{3a}=a_{30}\p_x^3 + a_{21}\p_x^2 \py + a_{12}\px
\py^2 + a_{03}\py^3 +
a_{20}\p_x^2+$$$$+a_{11}\px\py+a_{02}\py^2+a_{10}\px+a_{01}\py+a_{00},
$$

$$ A_{3p}=(p_1\px+p_2\py+p_3)(p_4 \p_x^2 +p_5 \px\py
+ p_6 \py^2 + $$$$+p_7 \px + p_8 \py + p_9)-l_3\py-l_{31}, $$

$$\tilde{p}_{i}=f^{-1}p_{i}\circ f, \ \
\tilde{a}_{i,j}=f^{-1}a_{i,j}\circ f,$$

$$\tilde{A}_{i}=f^{-1}A_{i}\circ f, \ \
i=2a,2p,3a,3p.$$

Above $A_{2p}=A_{2a}$, i.e. $A_{2p}$ and $A_{2a}$ are different
forms of the same operator - its initial form and its form after
the factorization with reminder. The same keeps true for $A_{3p}$
and $A_{3a}$, i.e. $A_{3p}=A_{3a}$.

\paragraph{Theorem 4.2 }
 For an operator of order 2, its reminder $l_2$ is
its {\bf invariant} under
 the equivalence transformation, i.e.
$$ \tilde{l}_{2}=l_{2}. $$ For an operator of order 3, its reminder
${l}_{3}$ is its {\bf invariant}, i.e. $$\tilde{l}_{3}=l_{3},$$
while reminder ${l}_{31}$ changes its form as follows:
 $$\tilde{l}_{31}=l_{31}+l_{3}\varphi_y. $$

$\blacktriangleright$ Indeed, for operator of order 2
$$
A_{2a}=A_{2p}-l_2,
$$
i.e.
$$
\tilde{A}_{2a}=f^{-1}A_{2a}\circ f= f^{-1} (A_{2p}-l_2)\circ
f=$$$$=\tilde{A}_{2p}-f^{-1}(l_2)\circ f=\tilde{A}_{2p}-l_2.
$$
For operator of order 3
$$
A_{3a}=A_{3p}-l_{3}\py-l_{31},
$$
i.e.
$$
\tilde{A}_{3a}=f^{-1}A_{3a}\circ f= f^{-1}
(A_{3p}-l_{3}\py-l_{31})\circ
f=$$$$=\tilde{A}_{3p}-f^{-1}(l_{31}+l_{3}\py)\circ
f=\tilde{A}_{3p}-l_{3}\py-l_{3}\varphi_y-l_{31}.\qed
$$

\paragraph{\bf Corollary 4.3: }
If $l_{3}=0$, then $l_{31}$ becomes invariant.\\

That is the reason why we call $l_{31}$ further {\bf
semi-invariant}.

\paragraph{\bf Corollary 4.4: }
If $l_{3} \neq 0$, it is always possible to choose some function
$f: \quad \tilde{l}_{31}= l_{3}\varphi_y+l_{31}=0$.\\

Notice that for second order operator, if its invariant
 $l_2=0$ then operator is factorizable while for third order
 operator two its invariants have to be equal to zero,
 $l_{3}=l_{31}=0$. On the other hand, if operator of third order
 is not factorizable we can always regard it as an operator with
only one non-zero invariant. Of course, all this is true for {\bf
each} distinct root of characteristic polynomial, so that one
expression, say, for $l_{3}$ will generate three invariants in
case of three distinct roots of corresponding polynomial.
Expressions for invariants $l_2$ and $l_{3}$ and also for
semi-invariant $l_{31}$ can be easily written out explicitly using
formulae given by BK-factorization (Section 2.1 and 2.2).\\

As it was show already, for an important particular case -
hyperbolic operator of second order in the form \be
\label{Lap}\p_x \p_y + a\p_x + b\p_y + c \ee - there exist two
Laplace invariants which coincide pairwise for equivalent
operators (Lemma 3.3). After rewriting hyperbolic operator in the
form \be \label{BK}\p_x^2 - \p_y^2 + a\p_x + b\p_y + c \ee by
appropriate change of variables, we can construct Laplace
invariants as a simple particular case from the formulae for
general invariants (see next
Section).\\

\paragraph{\bf Corollary 4.5: } Two hyperbolic second order
operators having the same normal form, (\ref{Lap}) or (\ref{BK}),
are equivalent {\bf iff} their general invariants coincide.

\subsection{Hierarchy of invariants}

As it was shown above, every general invariant is a function of a
distinct root $\o$ of the characteristic polynomial and each
distinct root provides one invariant. It means that for operator
of order $n$ we can get no more than $n$ different invariants.
Recollecting that BK-factorization in this case gives us one first
order operator and one operator of order $n-1$, let us put now
following question: are general invariants of operator of order
$n-1$ also invariants of
corresponding operator of order $n$?\\

Let regard, for instance, operator of order 3:
$$
{A}_{3a}={A}_{3p}={A}_{1}{A}_{2a}-l_{3}\py-l_{31}=
{A}_{1}({A}_{2p}-l_2)-l_{3}\py-l_{31}={A}_{1}{A}_{2p}-l_2{A}_{1}-l_{3}\py-l_{31}$$
and obviously
$$
\tilde{A}_{3a}={A}_{1}{A}_{2p}-l_2{A}_{1}-l_{3}\py-l_{31}=
\tilde{A}_{1}\tilde{A}_{2p}-l_2\tilde{A}_{1}-l_{3}\py-\tilde{l}_{31},
$$
i.e. $l_2$ is also invariant of operator ${A}_{3a}$. Let us notice
that general invariant $l_3=l_3(\o^{(3)})$ is a function of a
distinct root $\o^{(3)}$ of the polynomial
$$
\mathcal{P}_3(z)= a_{30}z^3+a_{21}z^2+a_{12}z+a_{03}
$$
while general invariant $l_2=l_2(\o^{(2)})$ is a function of a
distinct root $\o^{(2)}$ of the polynomial
$$
\mathcal{R}_2(z)=p_4 z^2 +p_5 z  + p_6
$$
with $p_4, \ p_5, \ p_6$ given by ({\it 3Pol}) for $\o=\o^{(3)}$.
In case of all distinct roots of both polynomials
$\mathcal{P}_3(z)$ and $\mathcal{R}_2(z)$, one will get maximal
number of invariants, namely 6 general invariants. Repeating the
procedure for an operator of order $n$, we get maximally $n!$
general invariants. In this way for operator of arbitrary order
$n$ we can construct the hierarchy of its general invariants
$$
l_n, l_{n-1}, ..., l_2
$$
and their explicit form is given by BK-factorization.\\

For instance, let us regard a third order hyperbolic operator in
the form

\be \label{ex4} C=a_{30}\p_x^3 + a_{21}\p_x^2 \py + a_{12}\px
\py^2 +a_{03}\p y^3+\mbox{terms of lower order} \ee

 with constant high order coefficients, i.e.
$a_{ij}=\const \ \forall \ i+j=3$ and all roots of characteristic
polynomial
$$
a_{30}\o^3+a_{21}\o^2+a_{12}\o+a_{03}=\mathcal{P}_3(\o)
$$
are distinct and real.  Then we can construct three simple
independent general invariants in following way. Notice first that
in this case high terms of (\ref{ex4}) can be written in the form
$$
(\a_1 \px + \b_1 \py)(\a_2 \px + \b_2 \py)(\a_3 \px + \b_3 \py)
$$
 for all non-proportional $\a_j, \b_i$ and after appropriate change
 of variables this expression can easily be reduced to

 $$
\px \py (\px+ \py).
 $$

Let us introduce notations

$$\pa_1=\pa_x,\ \
\pa_2=\pa_y,\ \ \pa_3=\pa_1+\pa_2=\p_t,$$

then all terms of the third and second order can be written out as

$$C_{ijk}=(\pa_i+a_i)(\pa_j+a_j)(\pa_k+a_k)=\pa_i\pa_j\pa_k+
a_k\pa_i\pa_j+a_j\pa_i\pa_k+a_i\pa_j\pa_k+$$
$$+
(\pa_j+a_j)(a_k)\pa_i+(\pa_i+a_i)(a_k)\pa_j+(\pa_i+a_i)(a_j)\pa_k+
(\pa_i+a_i)(\pa_j+a_j)(a_k)$$

 with $$a_{20}=a_2, \
a_{02}=a_1, \ a_{11}=a_1+a_2+a_3$$ and $c_{ijk}=C-C_{ijk}$ is an
operator of the first order which can be written out explicitly.
As it was shown above, coefficients of $c_{ijk}$ in front of first
derivatives {\bf are invariants} and therefore, any linear
combination of invariants is an invariant itself. These invariants
have the form:\\

for $c_{123}$ we have
$$
a_2a_3+a_1a_2+\py(a_3)+\px(a_2) -a_{10}, \ \
a_1a_3+a_1a_2+\px(a_3)+\px(a_2) -a_{01};
$$

for $c_{312}$ we have
$$
a_2a_3+a_1a_2+\p_t(a_2)+\px(a_2) -a_{10}, \ \
a_1a_3+a_1a_2+\p_t(a_1)+\px(a_2) -a_{01};
$$

for $c_{231}$ we have
$$
a_2a_3+a_1a_2+\py(a_3)+\py(a_1) -a_{10}, \ \
a_1a_3+a_1a_2+\p_t(a_1)+\py(a_1) -a_{01}.
$$

Direct calculation gives us three simplest general invariants of
the initial operator $C$:
$$l_{21}=a_{2,x}-a_{1,y}, \ l_{32}=a_{3,y}-a_{2,t}, \ l_{31}=a_{3,x}-a_{1,t}.$$

\paragraph{Proposition 4.6} General invariants $l_{21}, \ l_{32}, \ l_{31}$
are all equal to zero {\bf iff} operator $C$ is equivalent to an
operator
 \be \label{Prop}L=\pa_1\pa_2\pa_3+b_1\pa_1+b_2\pa_2+c,\ee
i.e. $\exists \ \mbox{function } \ f: \ \   f^{-1}C \circ f =L.$\\

 $\blacktriangleright$ Obviously
 $$
f^{-1}(\px \py \p_t )\circ f= (\px +(\log f)_x)(\py +(\log
f)_y)(\p_t +(\log f)_t)
 $$
for any smooth function $f$. Notice that it is the form of an
operator $C_{ijk}$ and introduce a function $f$ such that

$$
a_1=(\log f)_x, \ \ a_2= (\log f)_y, \ \ a_3= (\log f)_t.
$$

This system of equations on $f$ is over-determined
 and it has  solution $f_0$ {\bf iff}
 $$a_{2,x}-a_{1,y}=0, \ a_{3,y}-a_{2,t}=0,
\ a_{3,x}-a_{1,t}=0,$$ i.e. $l_{21}= l_{32}= l_{31}=0.$ \qed

$\blacktriangleleft$ Indeed, if $C$ is equivalent to (\ref{Prop}),
then $a_{20}=a_{02}=a_{11}=0$ and obviously
$l_{21}=l_{32}=l_{31}=0.$ \qed \\

At the end of this section let us notice that the reasoning above
can be carried out for the hyperbolic operator of order $n$ with
constant leading coefficients and analog of  Proposition 4.6 keeps
true with $n$ order terms as
$$
\p_1  \p_2 \p_3.... \p_n $$ with $$ \p_1=\px, \ \p_2=\py, \
\p_3=\px+\py, \ \p_4=\a_4 \px+ \b_4 \py, \ ... \ \p_n=\a_n \px+
\b_n \py
$$
and terms of order $n$ and $n-1$ can be written as
$$
C_{i_1,i_2,...,i_n}=(\p_1+a_1)(\p_2+a_2)...(\p_n+a_n)
$$

while corresponding linear general invariants will take form

$$
l_{ij}=\p_i(a_j)-\p_j(a_i).
$$

\section{Examples}
\subsection{Operator of order 2}
Let us regard first for simplicity  LPDO of the second order with
constant leading coefficients, i.e.
$$
a_{ij}=\const \quad \forall (i+j)=2
$$
and all roots of characteristic polynomial are distinct. Then
obviously any root $\o$ also does not depend on $x,y$ and
expressions for $p_{ij}$ can be simplified substantially. Let us
introduce notations
$$
a_{00}\o^0=\mathcal{P}_0(\o),
$$
$$
a_{10}\o^1+a_{01}=\mathcal{P}_1(\o),
$$
$$
a_{20}\o^2+a_{11}\o+a_{02}=\mathcal{P}_2(\o)
$$
and notice that now $\o$ and $ \mathcal{P}' _2(\o) \neq 0$ are
constants.\\

 Using formulae from Section 2.1 we get

\begin{eqnarray}\label{invar2}
p_1=1\nonumber\\
p_2=-\o \nonumber\\
p_3 =\frac{\mathcal{P}_1(\o)} {\mathcal{P}' _2(\o)}\nonumber\\
 p_4=a_{20}\nonumber\\
 p_5= a_{20} \o +a_{11}\nonumber\\
 p_6 =\frac{ (a_{20}\o+a_{11})a_{10}-a_{20}a_{01}
}{\mathcal{P}' _2(\o)}\nonumber
\end{eqnarray}

which yields to
$$
\mathcal{L}(p_6)=\frac{a_{20}\o+a_{11}}{\mathcal{P}' _2(\o)}
\mathcal{L}(a_{10})-\frac{a_{20}}{\mathcal{P}'
_2(\o)}\mathcal{L}(a_{01})
$$
and invariant $l_2$ takes form

\be \label{p7} l_2=
-\mathcal{L}(p_6)-p_3p_6+a_{00}=\frac{a_{20}\o+a_{11}}{\mathcal{P}'
_2(\o)} \mathcal{L}(a_{10})-\frac{a_{20}}{\mathcal{P}'
_2(\o)}\mathcal{L}(a_{01})+$$$$+\frac{\mathcal{P}_1(\o)}
{\mathcal{P}' _2(\o)}\frac{ (a_{20}\o+a_{11})a_{10}-a_{20}a_{01}
}{\mathcal{P}' _2(\o)}-a_{00}. \ee \\

\begin{itemize}
\item{} Let us regard hyperbolic operator in the form
\begin{equation}\label{ex1}
\partial _{xx} - \partial_{y y} + a_{10} \partial_{ x} +
a_{01} \partial_{ y} + a_{00},
\end{equation}
 i.e. $a_{20}=1, a_{11}=0, a_{02}=-1$ and $\o=\pm 1, \ \mathcal{L}=\px-\o \py.$
Then $l_2$ takes form
$$
l_2=a_{00}-\mathcal{L}(\frac{\o a_{10}-a_{01}}{2\o})-\frac{\o
a_{10}-a_{01}}{2\o} \frac{\o a_{10}+a_{01}}{2\o}
$$
which yields, for instance for the root $\o=1$, to
$$
l_2=a_{00}-\mathcal{L}(\frac{a_{10}-a_{01}}{2})-\frac{
a_{10}^2-a_{01}^2}{4}=a_{00}-(\px-
\py)(\frac{a_{10}-a_{01}}{2})-\frac{ a_{10}^2-a_{01}^2}{4}
$$
and after obvious change of variables in (\ref{ex1}) we get
finally first Laplace invariant $\hat{a}$
$$
l_2= c-\p_{\tilde{x}}a -ab = \hat{a},
$$
where
$$
a=\frac{a_{10}-a_{01}}{2}, \ \ b=\frac{a_{10}+a_{01}}{2}, \ \
c=a_{00}.
$$
Choice of the second root, $\o=-1$, gives us the second Laplace
invariant $\hat{b}$, i.e. Laplace invariants are particular cases
of the general invariant so that each Laplace invariant
corresponds to a special choice of $\o$.

\item{} Let us proceed analogously with an elliptic operator
\begin{equation}\label{ex2}
\partial _{xx} + \partial_{y y} + a_{10} \partial_{ x} +
a_{01} \partial_{ y} + a_{00},
\end{equation}
then $\o=\pm i, \ \mathcal{L}= \px-\o \py$ and
$$
l_2=a_{00}+(\px \mp i \py)(\frac{\pm a_{10}+a_{01}i}{2})+i\frac{
a_{10}^2+a_{01}^2}{4}
$$
where choice of upper signs corresponds to the choice of the root
$\o=i$ and choice of lower signs corresponds to $\o=-i$.
\end{itemize}

\subsection{Operator of order 3}

 Now let us regard  LPDO of the third
order with constant leading coefficients, i.e.
$$
a_{ij}=\const \quad \forall (i+j)=3
$$
with at least one  root distinct of characteristic polynomial
$$
a_{30}\o^3+a_{21}\o^2+a_{12}\o+a_{03}=\mathcal{P}_3(\o)
$$
and notice that now $\o$ and $ \mathcal{P}' _3(\o) \neq 0$ are
constants.\\

 Using formulae from Section 2.2 we get
\begin{eqnarray}
p_1=1\nonumber\\
p_2=-\o \nonumber\\
p_3 =
\frac{\mathcal{P}_2(\o)}{\mathcal{P}'_3(\o)}\nonumber\\
p_4=a_{30}\nonumber\\
p_5=a_{30} \o+a_{21}\nonumber\\
p_6=a_{30}\o^2+a_{21}\o+a_{12}\nonumber\\
 p_7=
\frac{a_{20}}{\mathcal{P}'_3(\o)}-\frac{a_{30}}
{\mathcal{P}'_3(\o)}\frac{\mathcal{P}_2(\o)}{\mathcal{P}'_3(\o)}\nonumber\\
 p_8= \frac{\o
a_{20}+a_{11}}{\mathcal{P}'_3(\o)}-\frac{\o a_{30}
+a_{21}}{\mathcal{P}'_3(\o)}\frac{\mathcal{P}_2(\o)}{\mathcal{P}'_3(\o)}\nonumber\\
p_9=a_{10}
-\frac{\mathcal{L}(a_{20})}{\mathcal{P}'_3(\o)}+\frac{a_{30}}
{\mathcal{P}'_3(\o)}\frac{\mathcal{L}(\mathcal{P}_2(\o))}{\mathcal{P}'_3(\o)}-
\frac{\mathcal{P}_2(\o)}{\mathcal{P}'_3(\o)}(
\frac{a_{20}}{\mathcal{P}'_3(\o)}-\frac{a_{30}}
{\mathcal{P}'_3(\o)}\frac{\mathcal{P}_2(\o)}{\mathcal{P}'_3(\o)})\nonumber
\end{eqnarray}
and $l_3, l_{31}$ are
$$-(p_1\px+p_2\py+p_3)p_8-p_2p_9+ a_{01} =l_3,$$
$$-(p_1\px+p_2\py+p_3)p_9+ a_{00}=l_{31}.$$
The formulae are still complicated and in order to show the use of
them let us regard here one simple example of an operator

\be\label{ex3} B=\p_x^2 \py + \px \py^2 +a_{11}\px\py
+a_{10}\px+a_{01}\py+a_{00}, \ee

with $a_{30}=a_{03}=a_{20}=a_{02}=0,\ \ a_{21}=a_{12}=1$. Then its
invariant
$$l_{3}=\px a_{11}-a_{01}$$
and semi-invariant $$l_{31}=\px a_{10}-a_{00}$$ have very simple
forms and gives us immediately a lot of information about the
properties of operators of the form (\ref{ex3}), for instance,
these operators are factorizable, i.e. has zero invariants
$l_3=l_{31}=0$, {\bf iff}
$$a_{11}=\int a_{01}dx + f_1(y), \ \ a_{10}= \int a_{00}dx +
f_2(y)
$$ with two arbitrary functions on $y$, $f_1(y)$ and $f_2(y)$.
Another interesting fact is that if coefficient $a_{11}=a_{11}(y)$
is function of one variable $y$, then $a_{00}$ is general
invariant and there definitely should be some nice geometrical
interpretation here, etc.

\section{Summary}

At the end of this Chapter we would like to notice following very
interesting fact - beginning with operator of order 4, maximal
number of general invariants is bigger then number of coefficients
of a given operator, $$\frac{(n+1)(n+2)}{2} < n! \ \ \forall
n>3.$$ It means that general invariants are dependent on each
other and it will be a challenging task to extract the subset of
independent general invariants,
i.e. basis in the finite space of general invariants.\\

 As to semi-invariants,
notice that an operator of arbitrary order $n$ can always be
rewritten in the form of factorization with reminder of the form
$$
l_n\px^k+l_{n,1}\px^{k-1}+...+l_{n,k-1}, \ \ k<n
$$
and exact expressions for all $l_i$ are provided by
BK-factorization procedure. The same reasoning as above will show
immediately that $l_n$ is always  general invariant, and each
$l_{n,k-i_0}$ is $i_0$-th semi-invariant, i.e. it becomes
invariant in case if $l_{n,k-i}=0, \ \forall i<i_0$.\\

In this paper, explicit formulae for $l_2, \ l_3, \ l_{31}$ are
given. Formulae for higher order operators can be obtained by pure
algebraic procedure described in \cite{bk2005} but they are too
tedious to be derived by hand, i.e. programm package for
symbolical computations is needed.\\

Already in the case of three variables, the factorization problem
of a corresponding operator and also constructing of its
invariants becomes more complicated, even for constant
coefficients. The reason of it is that in bivariate case we needed
just to factorize leading term polynomial which is always possible
over $\Complex$. It is not the case for more then 2 independent
variables where a counter-example is easily to find (see Ex.5),
i.e. there exist some non-trivial conditions to be found for
factorization of polynomials in more then two variables.

\section{Exercises for Chapter 5}

\paragraph{1.} Using this particular solution $u=-\frac{1}{(x+y)^2}$ of
the Liouville equation
$$(\log u)_{xy}+2u=0$$ find general solution of this equation.

{\bf Hint:} Liouville equation is invariant under the following
change of variables: $$\hat{x}=X(x),\quad \hat{y}=Y(y), \quad
\hat{u}(\hat{x},\hat{y})=\frac{u(x,y)}{X'(x)Y'(y)}.$$

\paragraph{2.} Prove that matrix (\ref{tildeAN}) is degenerate.

\paragraph{3.} Let Laplace invariants are equal, i.e.
$\hat{a}=\hat{b}$. Prove that initial operator $$ \p_x \p_y +
a\p_x + b\p_y + c$$ is equivalent to the operator $$ \p_x \p_y  +
c.$$

\paragraph{4.} Let in Lemma 3.8 function $w(x,y)$ is chosen as
$$w(x,y)= X_1(x)Y_1(y)+ X_2(x)Y_2(y).$$ Prove that $d_3=0.$

\paragraph{5.} Check that
$$
x^3+y^3+z^3-3xyz=(x+y+z)(x^2+y^2+z^2-xy-xz-zy)
$$
and prove that $ x^3+y^3+z^3$ is not divisible by a linear
polynomial $$\a x+ \b y+ \g z + \d$$ for any complex coefficients
$\a,\ \b,\ \g, \ \d$.
\section*{Acknowledgements}
 Author$^{1}$
acknowledges support of the Austrian Science Foundation (FWF)
under projects SFB F013/F1304.  Author$^{2}$ is very grateful to
RISC and J.Kepler University, Linz, for their hospitality during
preparing of this paper.

\end{document}